\documentclass[preprint,12pt]{elsarticle}

\usepackage{amssymb}
\usepackage{ulem}
\usepackage{color}

\journal{Journal of Magnetism and Magnetic Materials}

\begin{document}

\begin{frontmatter}



\title{Kondo Destruction in Heavy Fermion Quantum Criticality and the Photoemission Spectrum of YbRh$_2$Si$_2$}

\author[AA]{S. Paschen}
%\corauthref{Paschen}},\ead{paschen@ifp.tuwien.ac.at}
 \address[AA]{Institute of Solid State Physics, Vienna University of
 Technology, Wiedner~Hauptstr.~8-10, 1040 Vienna, Austria
}
 \author[BB]{S. Friedemann}
 \address[BB]{HH Wills Laboratory, University of Bristol, Bristol BS8 1TH, UK}
  \author[CC]{S. Wirth}
\author[CC]{F. Steglich}
 \address[CC]{Max Planck Institute for Chemical Physics of
Solids, N{\"o}thnitzer Str.~40, 01187~Dresden, Germany
}
\author[DD]{S. Kirchner}
 \address[DD]{Center for Correlated Matter, Zhejiang University, Hangzhou,  Zhejiang 310058, China}
\author[EE]{Q. Si}
 \address[EE]{Department of Physics and Astronomy, Rice University, Houston,
Texas 77005, USA
}

\begin{abstract}

Heavy fermion metals provide a prototype setting to study quantum criticality.
Experimentally, quantum critical points have been identified and studied in a
growing list of heavy fermion compounds. Theoretically, Kondo destruction has
provided a means to characterize a class of unconventional quantum critical
points that goes beyond the Landau framework of order-parameter fluctuations.
Among the prominent evidence for such local quantum criticality have been
measurements in YbRh$_2$Si$_2$. A rapid crossover is observed at finite
temperatures in the isothermal field dependence of the Hall coefficient and
other transport and thermodynamic quantities, which specifies a $T^*(B)$ line in
the temperature ($T$)-magnetic field ($B$) phase diagram. Here, we discuss what
happens when temperature is raised, by analyzing the ratio of the crossover
width to the crossover position. With this ratio approaching unity at $T \gtrsim
0.5$ K, YbRh$_2$Si$_2$ at zero magnetic field belongs to the quantum-critical
fluctuation regime, where the single-particle spectral function has significant
spectral weight at both the small and large Fermi surfaces. This implies that,
in this temperature range, any measurements sensitive to the Fermi surface will
also see a significant spectral weight at the large Fermi surface. The
angle-resolved photoemission spectroscopy (ARPES) experiments recently reported
for YbRh$_2$Si$_2$ at $T>1$ K are consistent with this expectation, and
therefore support the association of the $T^*(B)$ line with the physics of Kondo
destruction.

\end{abstract}


\end{frontmatter}

\section{Introduction}
\label{intro}
Quantum criticality is of extensive current interest in a variety of strongly
correlated electron systems \cite{qcnp.09,qcnp.12}. Heavy fermion metals have
emerged as prototype systems for quantum criticality
\cite{Coleman.05,hvl.2007,Si.2010}. The small energy scales of these materials
lead to increased tunability of the ground state by parameters such as pressure,
magnetic field or chemical substitution. This facilitates the realization of
quantum critical points (QCPs). Indeed, QCPs have been experimentally observed
in a considerable number of antiferromagnetic heavy fermion metals.

Heavy fermion metals have provided a setting to explore quantum criticality that
goes beyond the conventional Landau type of order-parameter fluctuations
\cite{Si.2013}. Theoretically, an unconventional type of quantum criticality has
been advanced, which is characterized by a critical destruction of the Kondo
effect \cite{Si.2001,Coleman.2001}. The Kondo destruction is manifested through
a jump of the Fermi surface {\it at zero temperature}, from small ({\it i.e.},
not incorporating the $f$-electrons) to large (involving the $f$-electrons).

Evidence for this local quantum criticality has come from a variety of
experiments. For example, inelastic neutron scattering experiments near the
Au-substitution induced QCP in CeCu$_{6}$ \cite{Schroder.00} and the
Pd-substitution induced QCP in UCu$_5$ \cite{Aronson} found the type of
dynamical scaling consistent with the theory. Moreover, de Haas-van Alphen
measurements \cite{Shishido.05} showed a jump from a small to a large Fermi
surface as pressure is increased through the antiferromagnetic QCP of CeRhIn$_5$
\cite{Park.06,Knebel.06}.

Measurements across the magnetic field-induced QCP in YbRh$_2$Si$_2$ have
revealed a rapid crossover in the isothermal field dependence of the Hall
coefficient and other magnetotransport and thermodynamic quantities
\cite{Paschen.04,Gegenwart.07,Friedemann.10,Nair.12}. These studies specify a
$T^*(B)$ line in the temperature-magnetic field phase diagram. Extrapolating
this crossover behavior towards lower temperatures has led to the conclusion
that, in the zero-temperature limit, the Hall coefficient and several other
transport and thermodynamic properties display a sudden jump across the critical
field.

In this paper, we discuss the implications of the above understandings for
ARPES, which can only be carried out at zero magnetic field and is presently
limited to relatively high temperatures. Towards this goal, we discuss what
happens to the isothermal crossover as the temperature is raised, by analyzing
the ratio of the crossover width to the crossover position. The resulting
temperature dependence of this ratio shows that, for $T \gtrsim 0.5$~K,
YbRh$_2$Si$_2$ at zero magnetic field fully belongs to the quantum-critical
fluctuation regime. In this regime, the single-particle spectral function has
significant spectral weight at both the small and the large Fermi surface. This
implies that ARPES measurements in this temperature range are expected to also
see a significant spectral weight at the large Fermi surface. The ARPES
experiments recently reported by Kummer {\it et al.} \cite{Kummer.15} on
YbRh$_2$Si$_2$ at $T \gtrsim 1$ K are consistent with the above expectation.

\section{Kondo effect and its critical destruction}
Magnetic heavy fermion metals contain a lattice of local moments that are
antiferromagnetically coupled to the spins of a conduction electron band.
Usually, the transition is between an antiferromagnetically ordered phase and
the paramagnetic metal phase.

Within the Landau framework, quantum criticality is described in
terms of the fluctuations of the antiferromagnetic order parameter.
The metallic nature would be manifested only through the presence of
Landau damping of the order parameter field by a decay into
particle-hole excitations. This is the picture of a spin-density-wave
(SDW) QCP \cite{Hertz.76,Millis.93,Moriya.95}. The effect of higher
order terms in the coupling between the order parameter and gapless
conduction electrons is the subject of continued theoretical interest.

\begin{figure}[t!]
\begin{center}
\includegraphics[width=0.45\textwidth]{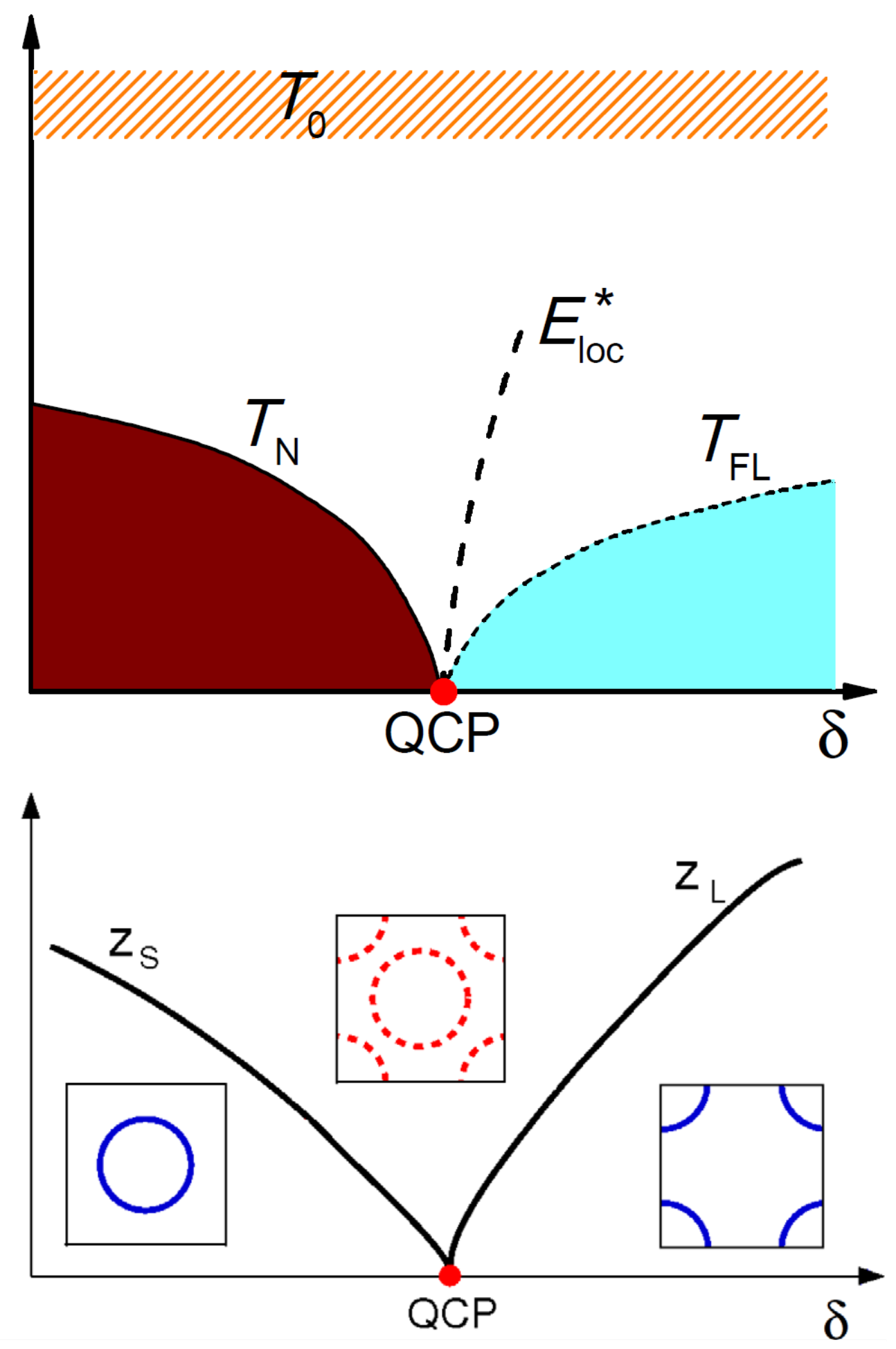}
\end{center}
\caption{Top: Kondo destruction at a heavy fermion QCP. Bottom: Small (left) and
large (right) Fermi surfaces (from \cite{Pfau.12}). Each Fermi surface is marked
by a solid line, which denotes a nonzero quasiparticle spectral weight $z_S$ or
$z_L$. In the quantum critical regime, while both $z_S$ and $z_L$ vanish,
low-energy single-particle excitations occur at both the small and large Fermi
surfaces and have a non-Fermi liquid form.}
\label{lqcp}
\end{figure}

However, the paramagnetic ground state of the heavy fermion metals
involves the lattice Kondo effect, with a nonzero amplitude for the
Kondo singlet, {\it i.e.,} entanglement between the local moments
and the spins of the conduction electrons. At a local QCP, this
Kondo-singlet amplitude goes to zero continuously as the QCP is
approached from the paramagnetic side \cite{Si.2001,Zhu.2003,Si.2014},
as the antiferromagnetic order sets in. This is illustrated
in Fig.~\ref{lqcp}, top panel. Here, $E_{loc}^*$ describes the energy
scale for the Kondo destruction. As the control parameter $\delta$
approaches $\delta_c$ from the paramagnetic side ($\delta > \delta_c$),
$E_{loc}^*$ vanishes at $\delta_c$, where the N\'eel order smoothly
sets in.

The critical destruction of the Kondo effect affects the scaling of the
order-parameter dynamics. The dynamical spin susceptibility at the QCP was shown
to have the following form:
\begin{eqnarray}
\chi({\bf q}, \omega ) =
\frac{1}{f({\bf q}) + A \,(-i\omega)^{\alpha} W(\omega/T)} \quad .
\label{chi-qw-T}
\end{eqnarray}
While the form itself has been derived analytically (within an
$\epsilon$-expansion) \cite{Si.2001,Si.2014}, the exponent $\alpha$ was
determined numerically. It was found to be close to 0.75 (ranging from 0.72 to
0.78 depending on the method of solution) in the case of an Ising-anisotropic
Kondo lattice \cite{Zhu.2003,Zhu.07,Glossop.07}.

In addition to the form of dynamical scaling and the extra energy scale
$E_{loc}^*$ vanishing at the QCP, the Kondo destruction is also manifested in
the evolution of the Fermi surface across the QCP \cite{Si.2014}. This is
illustrated in Fig.~\ref{lqcp}, bottom panel.
\begin{itemize}
\item For $\delta < \delta_c$ and {\it at sufficiently low temperatures}, the
Fermi surface is small and sharp. In other words, the single particle
excitations are propagating quasiparticles and gapless near the small Fermi
surface, but such excitations display a small energy gap at the large Fermi
surface.
\item For $\delta > \delta_c$, again at sufficiently low temperatures, the Fermi
surface is large and sharp. In other words, the single particle excitations are
propagating quasiparticles and gapless near the large Fermi surface; such
excitations have a small energy gap at the small Fermi surface.
\item In the crossover region, incoherent single-particle excitations exist at
both the small and the large Fermi surface. The quasiparticle residue for the
large Fermi surface, $z_L$, as well as its counterpart for the small Fermi
surface, $z_S$, vanish at zero energy and zero temperature; both depend on
energy and temperature in a power-law fashion. In other words, the
single-particle spectral weights for small but nonzero energies and temperatures
are nonzero at both the small and the large Fermi surface. At the same time, the
single-particle excitations assume a non-Fermi liquid form everywhere on the
Fermi surfaces (leaving no ``cold" portions of the Fermi surfaces).
\end{itemize}

\section{Evidence for Kondo destruction in quantum critical heavy
fermion metals}\label{signature}

\subsection{Dynamical scaling}
Inelastic neutron scattering experiments provide a means to measure the
dynamical spin susceptibility. Such experiments are challenging because they
require large single crystalline samples. Nontheless, results are available in
several quantum critical heavy fermion metals. The most prominent example is the
Ising-anisotropic CeCu$_{\rm 6-x}$Au$_{\rm x}$ at $x_c=0.1$, where the dynamical
spin susceptibility has been found \cite{Schroder.00} to have the form of
Eq.~(\ref{chi-qw-T}) and a critical exponent of about $0.75$. A similar form was
also found in the heavy fermion metal ${\rm UCu_{5-x}Pd_x}$ \cite{Aronson}.

\subsection{Fermi surface evolution}
\begin{figure*}[t!]
\centering\includegraphics*[width=0.8\linewidth]{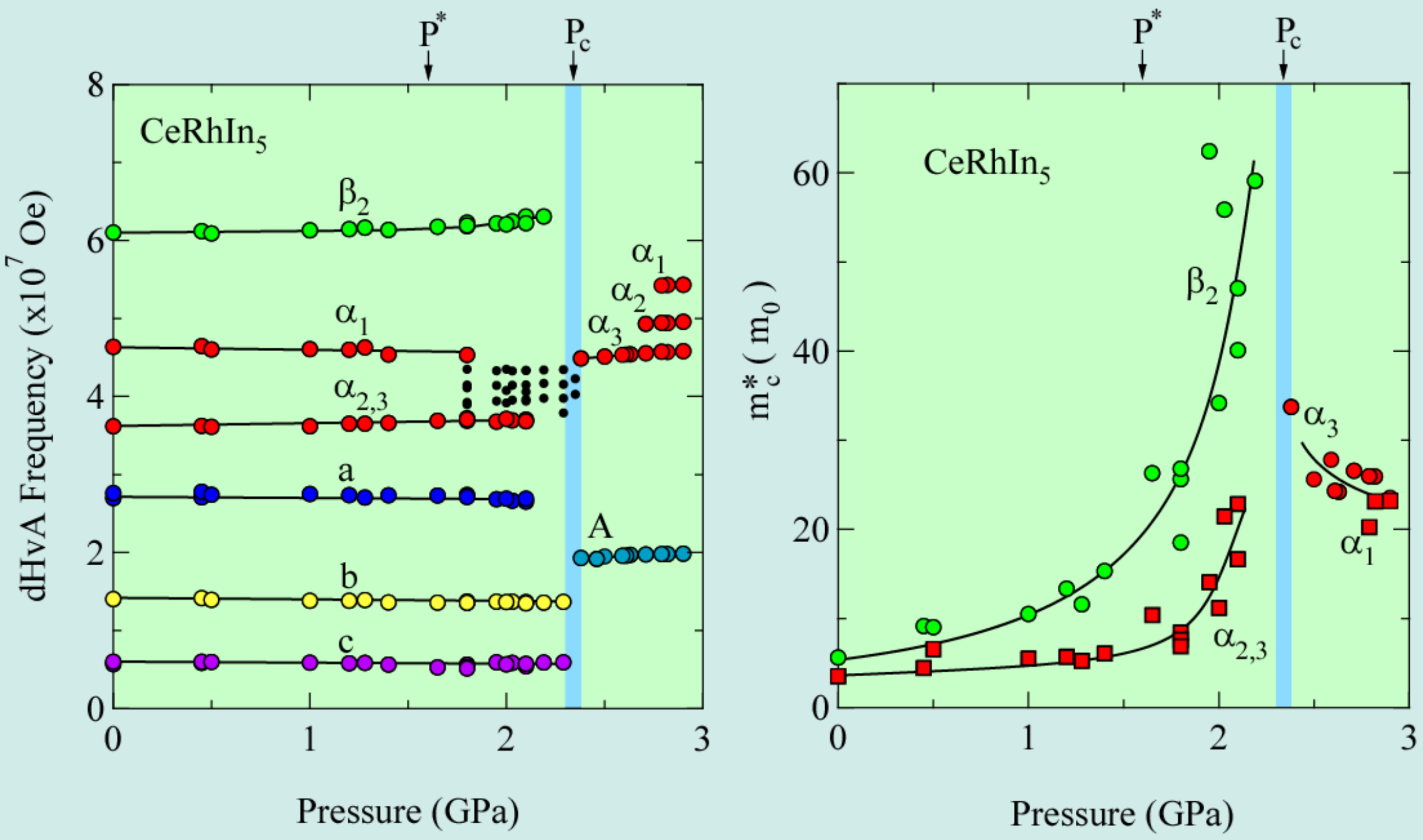}
\caption{Jump of Fermi surface across the critical pressure in CeRhIn$_5$,
observed by dHvA measurements (adapted from \cite{Shishido.05}).
}\label{dHvA}
\end{figure*}
CeRhIn$_5$ displays a continuous quantum phase transition as pressure is raised
across $p_c$ \cite{Park.06,Knebel.11}, at magnetic fields (of about 10 T) above
$H_{c2}$, the upper critical field for superconductivity. Quantum oscillations
have been observed by de Haas--van Alphen (dHvA) measurements at various
pressures at magnetic fields between 10 T and 17 T \cite{Shishido.05}. A jump of
the Fermi surface has been evidenced by the observation that the dHvA
frequencies undergo a sharp jump across $p_c$. The dHvA frequencies are
compatible with a small Fermi surface in the antiferromagnetically ordered state
at $p<p_c$, and with a large Fermi surface in the paramagnetic state at $p>p_c$.
The dHvA measurements also indicate that the cyclotron mass diverges at $p_c$,
providing evidence that the quasiparticle residues $z_L$ and $z_S$ indeed vanish
at the QCP, see Fig.~\ref{lqcp}, bottom.

\section{Extrapolating the isothermal crossover to lower temperatures}\label{crossover}

\begin{figure}[t!]
\begin{center}
\includegraphics[width=0.9\textwidth]{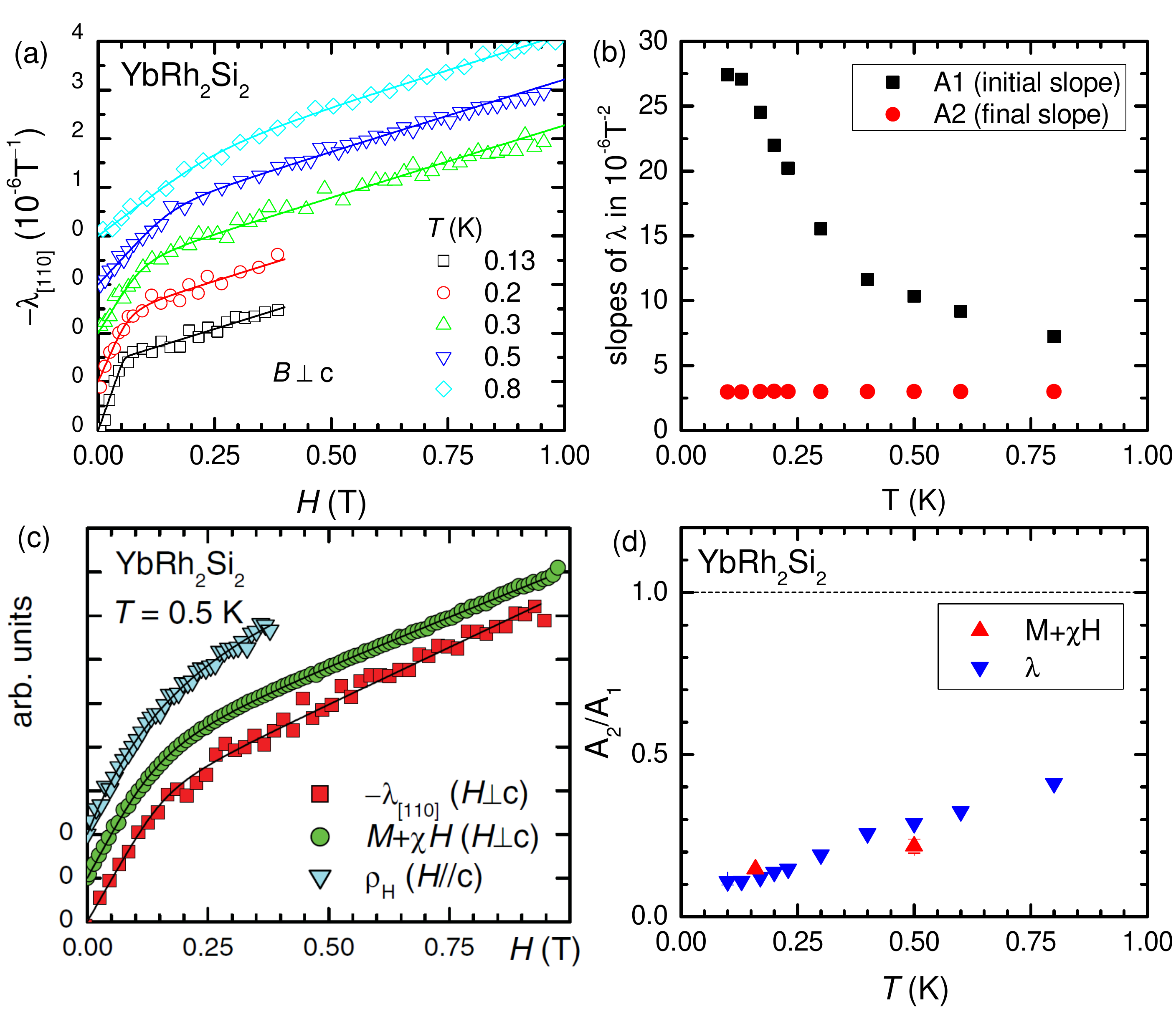}
\end{center}
\caption{Thermodynamic signatures of the $T^*(B)$ line. (a) Isothermal
magnetostriction $\lambda$ as a function of the magnetic field at selected 
temperatures \cite{Tanja}. (b) Initial slope $A_1$ for fields below $B^*(T)$,
and the final slope $A_2$ for fields above $B^*(T)$ as extracted from fits to
the magnetostriction. The difference between the two slopes, {\it i.e.} the
amplitude of the crossover, grows as temperature is reduced. (c) Isothermal
magnetostriction $\lambda$, $\tilde{M} = M + \chi H$ with the magnetisation $M$
and susceptibility $\chi$, and Hall resistivity $\rho_{\mathrm H}$ at 0.5\,K
\cite{Gegenwart.07}. The magnetization $M$ {\it vs.} $\rho_{\mathrm H}$ behaves
similarly as $\tilde{M}$ {\it vs.} $\rho_{\mathrm H}$ \cite{Gegenwart.07}. (d)
The ratio of the two slopes $A_2/A_1 \neq 1$ differing from 1 marks a finite
crossover amplitude in the magnetostriction and $\tilde{M}$ 
\cite{Gegenwart.07}.
}
\label{slopes}
\end{figure}

For YbRh$_2$Si$_2$, the isothermal crossovers as a function of magnetic field
$B$ have been studied between 0.02 K and 1 K. The pertinent measurements include
magnetotransport, Hall effect and magnetoresistance, as well as thermodynamic
properties, including magnetization and magnetostriction.

To draw conclusions about the nature of the QCP, the efforts have been directed
towards the evolution of the isothermal crossover behavior as temperature is
{\it lowered}. The lowest temperature of the studies is about 20 mK
\cite{Gegenwart.07}.  Importantly, as temperature is lowered, the full width at
half maximum (FWHM) of the isothermal crossover in all the measured properties
decreases \cite{Paschen.04,Gegenwart.07,Friedemann.10}. It extrapolates to zero
in the limit of zero temperature, consistent with a jump of the $T=0$ Hall
coefficient and related properties when the tuning parameter, the magnetic
field, crosses the QCP.

An example of the crossover behavior is illustrated in Fig.\ \ref{slopes}(a),
which shows the isothermal field dependence of the magnetostriction. For each
temperature, the field dependence can be fitted in terms of a crossover function
with an initial slope $A_1$, applicable to the regime of low fields prior to the
crossover, and a final slope $A_2$, applicable to high fields beyond the
crossover. The observed decrease of the slope ($\Delta A = A_2 - A_1 < 0$)
corresponds to a drop in the derivative $d\lambda/dB$ of the magnetostriction
isotherm. The amplitude of this drop {\it increases} as $T$ is lowered, thereby
extrapolating to a nonzero value in the zero-temperature limit. Together with
the vanishing width of the crossover, this implies a jump in $d\lambda/dB$.
Similar behavior has also been observed for the magnetization {\it vs.} $B$
(Fig.~\ref{slopes}(d)), and in the magnetotransport properties Hall coefficient
and magnetoresistance; in the latter two, the magnitude of the drop decreases as
$T$ is lowered, but extrapolates always to a non-zero value
\cite{Friedemann.11}. We take this as an  indication that the underlying
Fermi surface jump is a robust  effect and its manifestations in
magnetotransport are less  pronounced because of the interfering effects of
multiple bands that are present in YbRh$_2$Si$_2$ \cite{Friedemann.10}.

These studies lead to the conclusion that for $B=0 $ (and $B<B_c$ more
generally) and {\it at sufficiently low temperatures}, the Fermi surface is
small and sharp. For $B>B_c$, again at sufficiently low temperatures, the Fermi
surface is large and sharp. In the crossover region, incoherent single-particle
excitations exist at both the small and the large Fermi surface.

The important question is what happens at higher temperatures.

\begin{figure}[!ht]
\begin{center}
\includegraphics[width=0.98\textwidth]{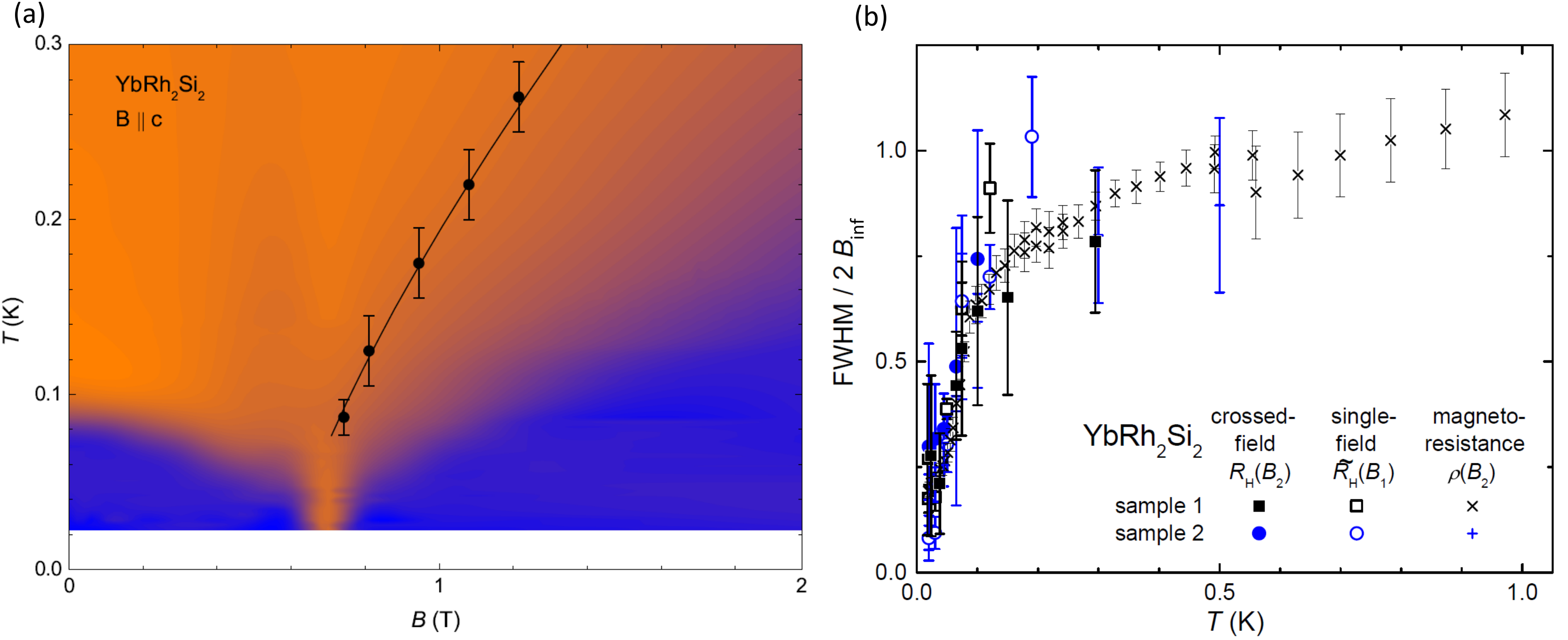}
\end{center}
\caption{(a) $T$--$B$ phase diagram. The color code indicates the exponent
$\varepsilon$ of the temperature dependent resistivity, {\it i.e.} $\rho (T) =
\rho_0 + c \, T^{\varepsilon}$ measured for $B || c$ \cite{Custers.03}. The
line indicates $T^*(B)$, as obtained from susceptibility measurements for $B
\perp c$ \cite{Tanja} and rescaled by a factor 11. (b) The ratio of the FWHM/2
of the isothermal crossover to $B_{\rm inf}$, where the magnetoresistance
exhibits an inflection point with respect to $B$. As temperature is raised to
about $0.5$ K and above, this ratio reaches $1$ within the error bars, implying
that $B=0$ is already part of the crossover regime.}
\label{HWHM_Bpeak}
\end{figure}

\section{Reaching up in temperature}\label{up}
One of the reasons for quantum criticality being of interest is that the unusual
excitations associated with the QCP ground state govern a large parameter regime
at nonzero temperatures. It is therefore of considerable interest to address how
a quantum critical system behaves as we reach {\it upwards} in temperature.

For YbRh$_2$Si$_2$, the isothermal crossover described in the previous section
can also be analyzed to infer the nature of the higher temperature portion of
the $T-B$ phase diagram. For reasons that will become clear, we are interested
in what happens at temperatures on the order of $1$ K and higher, at $B=0$. As
can be inferred from the $T$--$B$ phase diagram for $T \leq 0.3$ K, shown in
Fig.~\ref{HWHM_Bpeak}(a), for these higher temperatures the system obviously is
in the ``orange'' quantum-critical regime even at $B=0$. The same point can be
understood more quantitatively. In Fig.~\ref{HWHM_Bpeak}(b), we show the ratio
of the FWHM/2 associated with the isothermal crossover to $B_{\rm inf}$, defined
as the magnetic field at which the Hall coefficient and the magnetoresistance as
well as the field derivatives of the magnetization and the magnetostriction
exhibit an inflection point. As temperature is raised to about $0.5$ K and
above, this ratio reaches $1$ within error bars. The fact that this happens at
such low temperatures is tied up with the fact that $B^*(T)$ is relatively
small, which is in turn due to the small value of the critical field,
$B_c=B^*(T=0)$.

\section{Comments on the recent ARPES measurements}\label{comment}

\subsection{ARPES in YbRh$_2$Si$_2$}

Our considerations above show that for $T \gtrsim 0.5$ K the width of the
isothermal crossover is sufficiently large to make even $B=0$ to fall in the
crossover -- and quantum critical -- regime. This is consistent with the
defining property for Fig.\ \ref{HWHM_Bpeak}(a), namely that the electrical
resistivity at $B=0$ is linear in $T$ extending to very low temperatures on the
order of 0.1~K. In this temperature regime, the electronic Sommerfeld
coefficient, $\gamma \equiv C_{el}/T$ vs. $T$ at $B=0$ shows essentially the
same quantum critical behavior above $T_{\rm N}$ as its counterpart at $B=B_c$
\cite{Oeschler.08}.

Historically, observing heavy-fermion states using ARPES has been a challenge
because their relevant energy scale is very small. Nonetheless, there has been a
considerable amount of recent ARPES studies on heavy fermion metals, see below.
In particular, ARPES measurements with state-of-art resolution have been carried
out in recent years on YbRh$_2$Si$_2$ at $T>1$ K. For a particular part of the
Brillouin zone, finite spectral weight was observed for the temperature range
from $1$ K up to about $100$ K. It is expected that, at even higher
temperatures, the lattice Kondo effect should be suppressed and the Fermi
surface will be small. The lack of spectral weight at the small Fermi
surface for temperatures on the order of 100~K suggests that, in this ARPES
experiment, the matrix element is larger for the 4$f$-electron states than for
the conduction-electron states. It would be desirable to carry out ARPES
experiments to even higher temperatures in such a way that the small Fermi
surface can be observed.

Still, the ARPES experiments by Kummer {\it et al.} \cite{Kummer.15} at $B=0$
are instructive. They reached down to about $1$ K. Because at these elevated
temperature the system is in the crossover quantum critical regime already at
$B=0$, there should be low-energy electronic spectral weight at both the large
and the small Fermi surfaces, as discussed above. Thus, ARPES ought to see
spectral weight at the large Fermi surface, and it did.

Therefore, the ARPES measurements are consistent with a jump of the Fermi
surface as concluded from the low-temperature extrapolation of the isothermal
crossovers in magnetotransport and thermodynamic properties (section
\ref{crossover}).

\subsection{ARPES in YbCo$_2$Si$_2$ and other pertinent heavy
fermion compounds}

ARPES has become an indispensable tool in investigating the Fermi surface
topology of metals and has contributed significantly to our present
understanding of $d$-electron based materials in particular, as it gives direct
access to the momentum resolved single-electron spectral function. In the
context of $f$-electron based compounds like the heavy fermion metals with their
dynamically generated small energy scales, the utilization of
photoemission-based techniques has so far suffered from limited resolution in
momentum and energy. In recent years, significant progress in instrumentation
has led to an energy and momentum resolution close to the scales relevant for
these materials. As a result, a significant number of ARPES studies on heavy
fermion metals were recently reported. Besides YbRh$_2$Si$_2$
\cite{Kummer.15,Shen.12}, these include its trivalent counterpart YbCo$_2$Si$_2$
\cite{Guttler.14}, the antiferromagnets CeCu$_2$Ge$_2$~\cite{Kimura.15},
CeRhIn$_5$~\cite{Moore.02}, and the heavy fermion superconductors Ce$_2$CoIn$_8$
and Ce$_2$RhIn$_8$~\cite{Raj.05,Jiang.15}. Several groups have carried out ARPES
measurements on CeCoIn$_5$
\cite{Koitzsch.07,Koitzsch.08,Koitzsch.09,Jia.11,Koitzsch.13}. A 4$f$-derived
band persisting to comparably high temperatures has been reported in
\cite{Koitzsch.07}. But while the work reported in
\cite{Koitzsch.08,Koitzsch.09} seems to be more compatible with band-structure
calculations where the 4$f$-electron is treated as localized \cite{Koitzsch.09},
a more recent study identified a large Fermi surface~\cite{Jia.11} and discussed
the dependence of the result on photon energy. These results point to a moderate
to strong $k_z$ dependence of the hybridization in line with recent scanning
tunneling spectroscopy and optical conductivity measurements \cite{Koitzsch.13}.

The small energy scales present in the heavy fermion metals lead to good
tunability of the ground state with applied magnetic field or pressure. Making
use of this tunability within ARPES has so far been a challenge, see below. It
thus becomes pertinent to analyze ARPES data of different compounds at ambient
conditions within the general phase diagram of the heavy fermion compounds. For
YbRh$_2$Si$_2$, an ARPES study on its trivalent partner YbCo$_2$Si$_2$ has been
reported in Ref.~\cite{Guttler.14}. YbCo$_2$Si$_2$ possesses an
antiferromagnetic ground state formed by localized 4$f$ moments and displays an
only moderate mass enhancement as compared to YbRh$_2$Si$_2$. According to the
general phase diagram, when we get deep into the ordered phase, the Fermi
surface has to be small \cite{Yamamoto07}. Indeed, this is observed in
YbCo$_2$Si$_2$ together with a weak, band-like feature derived from a
crystalline electric field split 4$f$ state \cite{Guttler.14}.

\section{Discussion and Outlook}
\label{discussion}

Taken together with the anomalous dynamical scaling determined in the critically
Au-substituted CeCu$_6$ and the jump of the Fermi surface across the critical
pressure found by quantum oscillation experiments in CeRhIn$_5$, the isothermal
properties observed across the $T^*(B)$ line in the $T$--$B$ phase diagram of
YbRh$_2$Si$_2$ have been commonly recognized as the Kondo-destruction energy
scale expected in the local quantum critical description. As we have emphasized,
the $T^*(B)$ scale is not only observed by magnetotransport measurements, but is
also prominently reflected in thermodynamic quantities.

Unanimous evidence for, or against, a Fermi surface reconstruction across an
antiferromagnetic QCP could in principle be obtained by ARPES experiments
carried out at sufficiently low temperatures, as function of a non-thermal
control parameter crossing the QCP. Such experiments have for instance been
performed in the high-$T_c$ cuprates \cite{Dam.03} and the iron pnictides
\cite{Thi.11,Yos.11} as function of doping and/or chemical pressure.
Unfortunately, to detect the signature of the small Fermi surface in
YbRh$_2$Si$_2$ at very low temperatures upon reducing the magnetic field through
$B = B^*(T \! \rightarrow \! 0)$, the needed conditions are unfavorable for
ARPES: the temperature must be well below 0.5~K and the only known tuning
parameter is magnetic field. These challenging conditions might, however, be
achieved in future quasiparticle interference experiments using scanning
tunneling spectroscopy.

As for ARPES experiments on quantum critical heavy fermion systems, in addition
to tuning studies using chemical doping/pressure, uniaxial pressure experiments
might be a way forward. The challenges for this exciting emerging field,
however, remain formidable. To clearly resolve the heavy fermion excitations,
further breakthrough in energy resolution and in base temperature is needed.
Also the role of surface perfection/termination remains to be further
elucidated. ARPES experiments on MBE grown heavy fermion thin films may
contribute to further advancing the field.

We thank E. Abrahams, C. Geibel, P. Gegenwart, D. Vyalikh and G. Zwicknagl for
useful discussions. We are also grateful to Y. \={O}nuki for generating Fig.~2
from the data of Ref.~\cite{Shishido.05}, to P. Gegenwart for supplying the data
of Fig.~3(b) \cite{Gegenwart.07} and to J.\ Custers for providing part of
Fig.~4(a) \cite{Custers.03}. The work has been partially supported by the U.S.\
Army Research Office Grant No.\ W911NF-14-1-0497 and the European Research
Council Advanced Grant No.\ 227378 (SP), the German Research Foundation through
Forschergruppe 960 (SW and FS), the National Science Foundation of China Grant
No.\ 11474250 (SK), the U.S.\ Army Research Office Grant No.\ W911NF-14-1-0525,
NSF Grant No.\ DMR-1309531 and the Robert A.\ Welch foundation Grant No.\
C-1411 (QS).





\end{document}